\documentclass[12pt]{article}
\newcommand{\ket}[1]{|#1\rangle}

\newcommand{\phip}{\ket{\varphi^+}}

\newcommand{\unia}{\{\sigma_{0}, \sigma_{1}, i \sigma_{2}, \sigma_{3} \}}
\newcommand{\cn}{U^{CN}}

\begin{document}
\title{Task-Oriented  Maximally Entangled States}
\author{Pankaj Agrawal and B. Pradhan\footnote{email:
agrawal@iopb.res.in,bpradhan@iopb.res.in}\\
Institute of Physics \\ Sachivalaya Marg, Bhubaneswar, Orissa, India 751 005}
\maketitle
\begin{abstract}

  We introduce the notion of a task-oriented maximally entangled
  state (TMES). This notion depends on the tasks for which a quantum
  state is used as the resource. This concept may be more fruitful
  than that of a general maximally entangled state in the case 
  of a multipartite system. We illustrate this idea by
  giving an operational definition of maximally entangled states on
 the basis of communication tasks of teleportation and superdense
 coding. We also give examples and a procedure to obtain such TMESs for 
 {\em n}-qubit systems.

\end{abstract}

\vspace{1in}

%

\newpage

\section{Introduction}

       One of the sources of mysteries of quantum mechanics is
     the entanglement. Ever since this idea was introduced
     by Schrodinger in 1935 \cite{schr}, it has helped in unravelling the
     mystic of quantum mechanics. The entanglement of a quantum
     state of a system also allow us to carry out tasks, which would not be
     possible with a classical system. The entanglement properties
     of the quantum states are also responsible for the innate 
     nonlocality \cite{bell} of the quantum mechanical framework.
     Although this idea is quite old, 
     it is surprising that one still does not understand fully the 
     entanglement properties of a tripartite system in a pure
     state, or even a bipartite system in a mixed state \cite{pv}. The proper 
     characterization, quantification and classification of the entanglement 
     properties of a multipartite system \cite{foot1} is still a 
     developing field \cite{pv}.
       In this letter, we focus on systems in pure states. 

      A bipartite system in a pure state is the simplest system
   to examine and explore the entanglement properties. The nature 
   of entanglement of such a system is well understood.
   Such a system has only one bipartite quantum correlation which 
   can be characterized and quantified by the von 
   Neumann entropy \cite{pr}.  It is a suitable entanglement 
   measure \cite{vp,dhr} for such  states. This measure maps a 
   bipartite state to a real number. One can order the states according 
   to von Neumann entropy. 
     
    Because of the existence of a suitable entanglement measure, 
 there exists 
 the concept of a maximally entangled state. This concept is well defined
 for bipartite systems which are in pure states and is independent
 of the entanglement measure. The measure 
 von Neumann entropy
 is zero for the unentangled systems and one for the maximally
 entangled systems. The states of the maximally entangled bipartite
 systems are Bell states which are given below in (1) and (2). Beyond 
 a bipartite system, there is no consensus about what states might 
 be considered maximally entangled. There are numerous 
 suggestions in the literature \cite{pv} to characterize
 and quantify the entanglement features of a multipartite state. Some of
 these measures are quite hard to compute and it is not clear
 if they satisfy the criteria to be a suitable measure \cite{vp,dhr}. Without a
 suitable measure, what may be a maximally entangled state
 is not very clear. On the other hand, if one could identify the
 set of maximally entangled states, it could help in better
 understanding and classification of the multipartite
 entangled states.

   In this letter, we take a different approach to identify a maximally
entangled state. We suggest that for a multipartite system this
notion may not be universal. There may exist a global maxima
relative to a specific entanglement measure, but the state with 
such a property may not be most suitable for most tasks that 
one may envision. Therefore, we introduce
the notion of task-oriented maximally entangled
state (TMES). In this approach, there may not be an analog of Bell
states for multipartite systems (indeed such a search may not 
be very useful). Instead there may exist TMESs, that 
may be suitable to carry out a specific task 
{\em maximally}. For different
tasks, one may need different TMESs.  A task can be 
Bell inequalities or some equalities which use the correlations
of a quantum state or some communication or processing of information.
Some of the communication based tasks are teleportation \cite{bbcjpw},
superdense coding \cite{bw}, multi-receiver superdense coding \cite{pap},
quantum cryptography \cite{bb}, secret sharing \cite{hbb},
telecloning \cite{mjpv}, etc. This operational
approach may be more useful. We illustrate this concept by
  giving a definition of maximally entangled states on
 the basis of communication tasks of teleportation \cite{bbcjpw} 
and superdense
coding \cite{bw}. A quantum state might be considered maximally entangled 
on the basis of resources available to these communication protocols.
In this case, a TMES is the one which can help us to carry out the tasks
of teleportation and superdense coding maximally. On the basis of 
these  tasks, we can identify TMESs for a {\em n}-qubit system.
We can define maximality of these tasks as follows. For maximal 
teleportation, such an {\em n}-qubit state would allow us to teleport 
an unknown arbitrary $n \over 2$-qubit state, when {\em n} is even
and $(n-1) \over 2$-qubit state, when $n$ is odd. For maximal superdense
coding, a {\em n}-qubit state would allow us to transmit {\em n} classical
bits of information by sending ${n \over 2}$ qubits when {\em n} 
is even and ${(n + 1)\over 2}$ qubits when {\em n} is odd \cite{alt}. For 
example, when $ n = 4$, the quadripartite GHZ-state (given below in (3)) 
is not a TMES because one cannot teleport an unknown arbitrary
two-qubit state. Furthermore, although one can transmit
two cbits by sending one qubit and three cbits by sending two qubits,
one cannot transmit four cbits by sending two qubits from
Alice to Bob \cite{pap}. Therefore, this quadripartite state is not 
suitable for the maximal teleportation or maximal superdense coding.

     It is not surprising that a suitable entanglement measure, i.e. 
    von Neumann entropy exists 
     for bipartite states, since such states have only one bipartite
     correlation. However in the case of a multipartite entangled state, there
     exist multiple bipartite, tripartite and higher correlations. Therefore,
     one number may not be suitable to characterize such states. One may need
     a set of numbers to characterize such states. Another way to look at could
     be the following. All bipartite states can be characterized by just one parameter
     up to unary unitary transformation equivalency. However one needs more
     than one parameter to characterize a multipartite state. As an example,
     one needs six such real parameters to characterize a three-qubit 
     state \cite{eg}. It would
     appear to be unlikely that the complexity and richness of the state 
     of a multipartite system could be captured by just one number which 
     would characterized the state as an entanglement measure. One may
     need a set of numbers, a vector measure, to capture the multifaceted
     entanglement properties of such states.
        Therefore, if we insist on finding an ordered list of mulipartite entangled
     states, then it would depend on the correlations one is probing. In particular,
     there can be a number of different orderings which would depend on the
     task that we wish to perform using these states. Different tasks would
     depend on different quantum correlations which need to be maximal
     for that task. This naturally leads to the idea of TMESs.

\section{TMESs for even number of qubits}

         For a two-qubit system, the TMESs are well known. These are Bell
      states,
\begin{eqnarray}
\ket{\varphi^{\pm}} & =  & \frac{1}{\sqrt{2}}(\ket{00}\pm\ket{11})  \\
\ket{\psi^{\pm}} & =  & \frac{1}{\sqrt{2}}(\ket{01}\pm\ket{10}).
\end{eqnarray}

      Using these states, one can carry out the conventional teleportation
     and superdense coding. One can teleport one-qubit state perfectly.
     One can also transmit two cbits by sending one qubit. Therefore,
     with Alice and Bob one qubit each, both protocols can be 
     carried out maximally. Alice can convert these states
     into one another by applying unitary transformations $\unia$ on her
     qubit. Here $\sigma_0$ is a $2 \times 2$ unit matrix, and $\sigma_1,
     \sigma_2, \sigma_3$ are Pauli matrices. As a point  of 
     interest, these Bell states can be obtained from 
     the product states $\{ \ket{+}\ket{0}, \ket{-}\ket{0}, \ket{+}\ket{1},
     \ket{-}\ket{1} \}$ by applying the CNOT unitary operator $\cn$. As
     this operator acts on two qubits, it can entangle them.

      Let us now
     consider the case of larger number of qubits. Before considering
    a general {\em n}-qubit state, let us discuss the case of four-qubit
    states. There are a number of explicit states that have proposed
    as the maximally entangled state, as they seem to have
    some properties that may be desirable in a maximally entangled
    state. Some of these states
    are GHZ-state \cite{ghz} , cluster-state \cite{br,rb}, $\chi$-state \cite{yc}
   and H-S state \cite{hs},   

\begin{eqnarray}
  \ket{GHZ} & = & \frac{1}{\sqrt{2}}(\ket{0000} + \ket{1111}) \\
  \ket{\Omega} & = &  \frac{1}{2}(\ket{0000} + \ket{0110} +
                                            + \ket{1001} - \ket{1111})   \\
  \ket{\chi} & = &  \frac{1}{2 \sqrt{2}}(\ket{0000} - \ket{0011}
                    - \ket{0101} + \ket{0110} + \ket{1001} + \ket{1010} + \ket{1100}
                          + \ket{1111})\\
  \ket{HS} & = & \frac{1}{\sqrt{6}} (\ket{0011} + \ket{1100} +
                                           \omega(\ket{1010} + \ket{0101}) +
                                           \omega^2 (\ket{1001} + \ket{0110}) )
\end{eqnarray}

       Here $\omega = e^{2 \pi i \over 3}$.
       We would note that these are prototype states. By applying appropriate
    multi-unary \cite{trans} unitary transformations we can construct orthogonal states which would
     serve the same purpose. It is just like obtaining four Bell states
      from any one of the Bell states by applying appropriate unary 
      unitary transformations.
     These multi-unary transformations, by definition, act on qubits separately.
     There are various arguments that have been put forward to support
     the case of each state. Some of these are as follow. The
     GHZ-state has all one qubits in completely mixed state. The $\ket{\Omega}$
     state show certain features of entanglement which are persistent.
     The $\ket{\chi}$ state can be used for what we call maximal teleportation and
     superdense coding. The $\ket{HS}$ state has maximized two-qubit
     correlations. As we shall see later, these states can be TMESs, but for
     different tasks. But let us first construct TMESs for the tasks of teleportation
     and superdense coding.

     In case of
     even number of qubits, say 2{\em d}, one can teleport at most an arbitrary
     unknown {\em d}-qubit state. Easiest way to do it is to use {\em d} Bell states.
      Then each qubit of the {\em d}-qubit state can be teleported using
      one of the Bell states. This means that for the larger number of qubits,
     one could take a direct product of Bell states as the quantum resource.
     However, as these resource states are not genuinely multipartite
     entangled states, this resource is not very interesting. However, it turns
     that by a multinary \cite{trans} unitary transformation, one could convert
     these unentangled states into entangled states. Since the teleportation
     protocol is not affected by these unitary transformations, these states
     with genuine multipartite entanglement also serve the
     purpose of teleporting a {\em d}-qubit state. We can see this
     as follows. Suppose we wish to teleport an arbitrary unknown
     {\em n}-qubit state $\ket{\psi}$ using the {\em m}-qubit state $\ket{R}$
     as a quantum resource ($m \geq 2 n$). If the teleportation is successful, 
     then we would be able to write,

\begin{eqnarray}
       \ket{\psi}_{a_1 a_2.....a_n} \ket{R}_{b_1 b_2.....b_m} & = &
        {1 \over 2^{n}}  \sum_{i=1}^{2^{2n}} \ket{{\cal O}^i}_{a_1 a_2.....a_n b_1 b_2....b_{m-n}} {V^{i}}^{\dagger}_{b_{m-n+1} 
b_{m-n+2}....b_m} \ket{\psi}_{b_{m-n+1} 
b_{m-n+2}....b_m}. 
\end{eqnarray}

     Here subscripts are particle labels.
     $\ket{{\cal O}^i}$ are a set of orthogonal states
     and $V^{i}$ are unitary operators.  
     Alice would need to send $2n$ cbits of information to Bob to teleport
     a $n$-qubit state. 
     If we apply an unitary transformation $ I_{a_1 a_2......a_n} \otimes
     U_{b_1........b_{m-n}}$ on the both side of the equation, then this
      transformation would convert the set of orthogonal states $\ket{{\cal O}^i}$
      to another such a set. So the teleportation would still be possible, but
      with a different quantum resource state, 
     $ U_{b_1........b_{m-n}} \ket{R}_{b_1 b_2.....b_m} $. Therefore, 
     if we know a quantum resource state (e.g., a product of Bell states)
     that would allow the teleportation of an
      arbitrary unknown $n$-qubit state, then we can find another
      resource state by applying appropriate unitary transformation 
     to it \cite{czg}. Alice can apply this unitary transformation.
     (Instead of Alice, Bob can also do it.) Interestingly,  
     these entangled states can also be used for maximal superdense
     coding for even number of qubits. This may not be surprising 
     because of the close relationship
     between the two tasks. From the procedure, it is also clear
     that if a resource state can be used to teleport a $n$-qubit state,
     it can also be used to teleport any $p$-qubit state such that
     $p < n$. We illustrate the procedure for the four-qubit systems.

        With a four-qubit entangled state, maximal teleportation
     will be that of an arbitrary and unknown two-qubit state,

\begin{equation}
\ket{\psi}_{ab} = \alpha \ket{00}_{ab}+\beta \ket{01}_{ab}+
                             \gamma \ket{10}_{ab}+ \delta \ket{11}_{ab}.
\end{equation}

      Here $\alpha, \beta, \gamma$ and $\delta$ are complex numbers.
      As noted above one could use the direct product of two Bell states
     as a resource. Since there are four Bell states, there are sixteen possibilities.
     Now we could apply binary unitary operator on two particles
      of these two states. For concreteness, let us take a situation where
     Alice and Bob share two copies of the $\phip$ states. One copy has
     particles 1 and 2, and the second copy has particles 3 and 4. Let
     Alice has particles 1 and 3, and Bob has particles 2 and 4. Here,
     particles 1 \& 3 and particles 2 \& 4 are not entangled. Alice can now
     apply the {\em CNOT} operation on her qubits. This will entangle
     her qubits and one would obtain a genuine quadripartite entangled
     state. This state is actually an example of cluster state,

\begin{eqnarray}
    U^{CN}_{13} \phip_{12} \phip_{34} & = & \frac{1}{2}\cn_{13} (\ket{0000} + \ket{0011}
                                     + \ket{1100} + \ket{1111})_{1234} \nonumber \\
       & = & \frac{1}{2}(\ket{0000} + \ket{0011}
                                                     + \ket{1110} + \ket{1101})_{1234}.
\end{eqnarray} 
        This is a cluster sate and one can explicitly check that one can indeed
     use it to teleport two-qubit unknown state and also for maximal
     super dense coding \cite{pap}.

       Using a different unitary operator, one can generate $\ket{\chi}$ states
     also. The $\cn$ operator has the following representation in the
     computational basis,
\begin{equation}         
        \cn = \left(
             \begin{array}{cc}
                   \sigma_{0}   &  0 \\
                       0   &   \sigma_{1}
\end{array} \right).
\end{equation}
   Replacing $\sigma_{1}$
   by $\sigma_{3}$ in this matrix will give us the cluster state given in (4).
   The binary unitary operator
    that is needed to convert $\phip \phip$ to the $\ket{\chi}$ state
    is,
 \begin{equation}         
        U^{\chi} = {1 \over \sqrt{2}} \left(
             \begin{array}{cc}
                   \sigma_{3}   &  \sigma_{1} \\
                       i \sigma_{2}   &   \sigma_{0}
\end{array} \right).
\end{equation}

The state  $\ket{\Omega}$ and the state $\ket{\chi}$ are TMESs
     from the point of view of the tasks of teleportation and
     superdense coding. The GHZ-state and HS-state are {\em not} TMESs
     for these tasks, as we can check that these states are not suitable
     for maximal teleportation and superdense coding. The transformations
    that can convert two Bell states to these stats are also not
    unitary, as it should. The HS-state
     can be a TMES for the task of telecloning. This protocol requires
     maximized and equal bipartite correlations \cite{mjpv}. The 
     HS-state is designed that way. For quantum secret sharing,
     $\ket{GHZ}, \ket{\Omega}$, and $\ket{\chi}$ states can be TMESs,
     but the $\ket{HS}$ state cannot be.

        There are many different sets of TMESs from the point of view
    of maximal teleportation and superdense coding protocols. We have
   already seen two such sets. One set is that of sixteen orthogonal cluster 
   states, other is that of sixteen orthogonal $\chi$-state. These sets
   are obtained from the original states by applying multi-unary unitary
   transformations. One can generate different sets by applying 
   different binary unitary operator on the product of two Bell states.
   In fact one can construct sixteen linearly independent binary
   unitary operators that act on two qubits. So we could have sixteen 
   independent such sets. 
      One interesting set of sixteen linearly independent operators including
    CNOT unitary operator, $\cn$ is:

 \[         
        \Gamma_{1} = \left(
             \begin{array}{cc}
                   \sigma_{0}   &  0 \\
                       0 &   \sigma_{1}
\end{array} \right),
        \Gamma_{2} = \left(
             \begin{array}{cc}
                   \sigma_{1}   &  0 \\
                        0 &   \sigma_{2}
\end{array} \right),
        \Gamma_{3} = \left(
             \begin{array}{cc}
                   \sigma_{2}   &  0 \\
                       0  &   \sigma_{3}
\end{array} \right),
        \Gamma_{4} = \left(
             \begin{array}{cc}
                   \sigma_{3}   &  0 \\
                       0  &   \sigma_{0}
\end{array} \right), 
\]
\[
        \Gamma_{5} = \left(
             \begin{array}{cc}
                   \sigma_{0}   &  0 \\
                      0  &   -\sigma_{1}
\end{array} \right),
        \Gamma_{6} = \left(
             \begin{array}{cc}
                   \sigma_{1}   &  0 \\
                      0  &   -\sigma_{2}
\end{array} \right),
        \Gamma_{7} = \left(
             \begin{array}{cc}
                   \sigma_{2}   &  0 \\
                      0  &   -\sigma_{3}
\end{array} \right),
        \Gamma_{8} = \left(
             \begin{array}{cc}
                   \sigma_{3}   &  0 \\
                      0  &   -\sigma_{0}
\end{array} \right), 
\]
\[
        \Gamma_{9} = \left(
             \begin{array}{cc}
                   0 & \sigma_{0}  \\
                 \sigma_{1} & 0
\end{array} \right),
        \Gamma_{10} = \left(
             \begin{array}{cc}
                  0  &   \sigma_{1}  \\
                 \sigma_{2} &  0
\end{array} \right),
        \Gamma_{11} = \left(
             \begin{array}{cc}
                 0  &    \sigma_{2}  \\
               \sigma_{3} &  0
\end{array} \right),
        \Gamma_{12} = \left(
             \begin{array}{cc}
                 0  &    \sigma_{3} \\
                  \sigma_{0}& 0
\end{array} \right),
\]
\[
        \Gamma_{13} = \left(
             \begin{array}{cc}
                   0 & \sigma_{0}  \\
                   -\sigma_{1} & 0
\end{array} \right),
        \Gamma_{14} = \left(
             \begin{array}{cc}
                  0  &   \sigma_{1}  \\
                 -\sigma_{2} &  0
\end{array} \right),
        \Gamma_{15} = \left(
             \begin{array}{cc}
                 0  &    \sigma_{2}  \\
             -\sigma_{3} &  0
\end{array} \right),
        \Gamma_{16} = \left(
             \begin{array}{cc}
                 0  &    \sigma_{3} \\
                 - \sigma_{0}& 0
\end{array} \right). 
\]

     If we wish all above matrices to be real, then we can replace
    $\sigma_2$ by $i \sigma_2$ . These matrices are the representation
    of the operators in the computational basis. Interestingly, all of
    these operators will generate entangled states from the product
    of two Bell states.
      Similarly, one can obtain other sets including  CNOT like
    operators controlled-Y and controlled-Z, 

\begin{equation}
        U_{Y} = \left(
             \begin{array}{cc}
                 \sigma_{0} & 0 \\
                    0 &  \sigma_{2}
\end{array} \right),
        U_{Z} = \left(
             \begin{array}{cc}
                  \sigma_{0} & 0\\
                 0  &\sigma_{3}
\end{array} \right).
\end{equation}

      One can obtain other sets by applying unitary transformation
    on the above matrices.
       To obtain TMESs for the six-qubit systems and beyond, one can apply
    CNOT unitary operator on two pairs of qubits. For example, if Alice
    has qubits 1,3 and 5, while Bob has the qubits 2,4 and 6, then
    Alice can apply $\cn_{13}$ and $\cn_{35}$ to obtain a TMES.
    As noted above this TMES is not unique. One could apply any
    other set of binary or even a multinary unitary operator. 
    By applying such linearly independent unitary operators, 
    one can generate all sets. 

   This process can be extended to any 2{\em d}-qubit state.
    In particular, one could apply CNOT operations, or any other
    binary unitary operations suggested above,  on
   suitable $d-1$ pairs of qubits. Or, one may choose to apply
   a multinary unitary operator on any other subsets or on
   all $d$ qubits. This procedure
   will give us genuinely entangled $n$-qubit states ($n$ even)
   which are TMESs. One way to obtain an independent set of
   such multinary operators is to extend the procedure that we have
   used to obtain the set $\{ \Gamma_a \}$ that act on
   two-qubits using the set $\{ \sigma_a \}$ that act on one-qubit.
   This procedure allows us to construct the matrices for $(d+1)$-qubits
   from those that act on $d$-qubits. For $d = 1$ and $2$, we have
   given these matrices above. Let the independent set
   of the unitary operators that act on $d$-qubits be $\{ \gamma_a \}$,
   where $a = 1, 2,.....2^{2d}$. Then for the $(d+1)$-qubits, we
   can construct $2^{2d+2}$ linearly independent unitary matrices
   $\Sigma_{a}$ as follow,

\begin{scriptsize}
\[
        \Sigma_{1} = \left(
             \begin{array}{cc}
                   \gamma_{1}   &  0 \\
                       0 &   \gamma_{2}
\end{array} \right), 
        \Sigma_{2} = \left(
             \begin{array}{cc}
                   \gamma_{2}   &  0 \\
                        0 &   \gamma_{3}
\end{array} \right), ......., 
        \Sigma_{2^{2d}-1} = \left(
             \begin{array}{cc}
                   \gamma_{2^{2d} - 1}   &  0 \\
                       0  &   \gamma_{2^{2d}}
\end{array} \right), 
        \Sigma_{2^{2d}} = \left(
             \begin{array}{cc}
                   \gamma_{2^{2d}}   &  0 \\
                       0  &   \gamma_{1}
\end{array} \right),  
\]
\[
        \Sigma_{2^{2d}+1} = \left(
             \begin{array}{cc}
                   \gamma_{1}   &  0 \\
                      0  &   -\gamma_{2}
\end{array} \right), 
        \Sigma_{2^{2d}+2} = \left(
             \begin{array}{cc}
                   \gamma_{2}   &  0 \\
                      0  &   -\gamma_{3}
\end{array} \right), ......., 
        \Sigma_{2^{2d+1}-1} = \left(
             \begin{array}{cc}
                   \gamma_{2^{2d} - 1}   &  0 \\
                      0  &   -\gamma_{2^{2d}}
\end{array} \right), 
        \Sigma_{2^{2d+1}} = \left(
             \begin{array}{cc}
                   \gamma_{2^{2d}}   &  0 \\
                      0  &   -\gamma_{1}
\end{array} \right), 
\]
\[
        \Sigma_{2^{2d+1}+1} = \left(
             \begin{array}{cc}
                   0 & \gamma_{1}  \\
                 \gamma_{2} & 0
\end{array} \right), 
        \Sigma_{2^{2d+1}+2} = \left(
             \begin{array}{cc}
                  0  &   \gamma_{2}  \\
                 \gamma_{3} &  0
\end{array} \right), ........, 
        \Sigma_{3\times2^{2d}-1} = \left(
             \begin{array}{cc}
                 0  &    \gamma_{2^{2d}-1}  \\
               \gamma_{2^{2d}} &  0
\end{array} \right), 
        \Sigma_{3\times2^{2d}} = \left(
             \begin{array}{cc}
                 0  &    \gamma_{2^{2d}} \\
                  \gamma_{1}& 0
\end{array} \right), 
\]
\[
        \Sigma_{3\times2^{2d}+1} = \left(
             \begin{array}{cc}
                   0 & \gamma_{1}  \\
                 -\gamma_{2} & 0
\end{array} \right), 
        \Sigma_{3\times2^{2d}+2} = \left(
             \begin{array}{cc}
                  0  &   \gamma_{2}  \\
                 -\gamma_{3} &  0
\end{array} \right), .......,
        \Sigma_{2^{2d+2}-1} = \left(
             \begin{array}{cc}
                 0  &    \gamma_{2^{2d}-1}  \\
              - \gamma_{2^{2d}} &  0
\end{array} \right),
        \Sigma_{2^{2d+2}} = \left(
             \begin{array}{cc}
                 0  &    \gamma_{2^{2d}} \\
                  -\gamma_{1}& 0
\end{array} \right). 
\]
\end{scriptsize}

       It is our conjecture that using this procedure, one can,
    in principle, generate all suitable states that can be quantum
    resource in teleporting an arbitrary and unknown $d$-qubit state
    and for superdense coding.

\section{TMESs for odd number of qubits}

     For the systems with odd number of qubits, first non-trivial system
    is that of three qubits. None of the three-qubit states cannot be used to 
    teleport an unknown arbitrary two-qubit state \cite{ap2}. This is
   also clear from the fact that the construction of this state requires 
   only one Bell state, as we shall see below. Furthermore
    since the Hilbert space of a three-qubit system is only eight-dimensional, 
    one cannot transmit four cbits by transmitting two qubits. At most 
    three cbits could be transmitted.
   To obtain an TMES for  2{\em d} + 1  qubits, one can start with
   the tensor product of {\em d} Bell states and a computational basis
   state. In the case of three-qubits system, e.g., we could have,
\begin{eqnarray}
    U^{CN}_{13} \phip_{12} \ket{0}_{3} & = & \frac{1}{\sqrt{2}}U^{CN}_{13} 
                                  (\ket{000}_{123} + \ket{110}_{123})    \nonumber \\
       & = & \frac{1}{\sqrt{2}}(\ket{000}_{123}  + \ket{111}_{123} ).
\end{eqnarray} 

         This is a GHZ state. For a three-qubit systems it is a TMES for the
    tasks of superdense coding and teleportation. We note there exist
    a subclass of W-sates \cite{ap} which can also be used for the teleportation,

\begin{equation}
 \ket{W_{n}} = \frac{1}{\sqrt{2+2n}}(\ket{100}+\sqrt{n}\ket{010}+
\sqrt{n+1}\ket{001}).
\end{equation}

    Since this state can be used for the maximal teleportation, one should
   be able to obtain it after applying suitable multinary unitary 
   transformations on $\phip \ket{0}$. For example this transformation
   for the $\ket{W_2}$ state is,

\begin{equation}
        U^{W_2} = \left(
             \begin{array}{cccc}
                   0 & 1 \over \sqrt{2} & 1 \over \sqrt{2}  &  0 \\
                   0 & 0 &  0  &  1 \\
                   1  & 0  &  0  &  0\\
                    0 & 1 \over \sqrt{2}  & -1 \over \sqrt{2}    & 0
\end{array} \right).
\end{equation}

     This state is also suitable for transmitting three cbits
   by sending two qubits, but not for all $2-1$ partitions of three qubits. 
   Since our definition of the maximal superdense coding does not have any
   requirement for partitions, this state is
   also a TMES.  We note that the product state $\phip \ket{0}$ can be
   used for maximal superdense coding also. 

     Let us now consider a system of five qubits. For such a system
  TMES would not be a GHZ state \cite{pap}. However, we can obtain
  TMESs by applying appropriate unitary transformations on the product
   state of two Bell states and a computational basis state. One way
  is,
\begin{eqnarray}
    U^{CN}_{13} U^{CN}_{35}\phip_{12} \phip_{34} \ket{0}_{5} & = & 
                     \frac{1}{2}
                     U^{CN}_{13} U^{CN}_{35} (\ket{00000} + \ket{00110}
                                   + \ket{11000} + \ket{11110})_{12345}  \nonumber \\
       & =  & U^{CN}_{35} \frac{1}{2}(\ket{00000} + \ket{00110}
                            + \ket{11100} + \ket{11010})_{12345}  \nonumber \\
       & = &  \frac{1}{2}(\ket{00000} + \ket{00111}
                                                               + \ket{11101} + \ket{11010})_{12345} .
\end{eqnarray} 

     This state is a TMES for a five-qubit system corresponding to the tasks
   of teleportation and superdense coding. This is because it can be 
   used for maximal superdense coding and teleportation. One can teleport
   an unknown arbitrary two-qubit state and transmit five cbits
   by sending three qubits. This is also an example of cluster state for five qubits.
   As discussed in the case of even number of qubit system, this state
   is not unique. By applying multi-unary unitary transformations on three
   qubits, one can obtain thirty-two orthogonal states, which will serve
   the same purpose. Furthermore, by applying linearly independent
   multinary unitary transformations, given in the last section,
   one can obtain linearly independent sets of TMESs.  
     By maximizing negativity numerically,  Brown et al \cite{bssb} have obtained a 
   ``highly entangled'' five-qubit state. Their state is a TMES for ours tasks.
    One can obtain it by applying an appropriate multinary unitary transformation.
    This state does not have all maximized bipartite correlations.
                       
       One can generalize this procedure to any odd number of qubit system.
    So, for a 2{\em d} + 1 qubit system one has to consider the product
    state of {\em d} Bell states and a one-qubit state and apply a suitable
    multinary unitary transformation. This will make the 2$d$ + 1 qubits
    genuinely entangled. This can serve as the resource for the 
   maximal teleportation and superdense coding. In the last section
    we have given a more detailed procedure and sets of unitary matrices 
   that can be used for the transformations.

\section{Conclusions}

     We have argued that the universal maximally entangled
   state in the case of multipartite systems may not be generally
   useful, even if
   such a state could be identified. Instead 
   the notion of a task-oriented maximally entangled
   state (TMES) may be more fruitful. This makes the notion of 
   maximally entangled states task-dependent. For different tasks,
   different states may allow us to carry out the task maximally.
   We have illustrated this idea
   on the basis of the communication tasks of teleportation 
   and superdense coding. To this end, we have given strategy to
   obtain instances of the TMESs corresponding
   to these tasks. We conjecture that using our procedure one
   would be able to obtain all states that may be suitable for
   the teleportation and superdense coding. Similarly, for other tasks, 
   one may obtain the suitable TMESs.


\begin{thebibliography}{99}
\bibitem{alt} There is an alternate way to define maximal superdense
   coding. Here the definition remains the same for even number of
   qubits. In the case of odd number of qubits, $n-1$ cbits are
   transmitted by sending $({n-1 \over 2})$ qubits. If we adopt this definition,
   then all the states that can do maximal teleportation can also do
   maximal superdense coding.
\bibitem{trans} The unary transformations act on only one qubit. 
   The multi-unary transformations act separately on qubits of a multi-qubit
   system; they are direct product of unary transformations.
   The multinary transformations acts on multiple qubits at the same time.
    These are entangled transformations. They can change the
    entanglement properties of the system.
\end{thebibliography}
\end{document}